\documentclass{PoS}

\usepackage{fancybox}
\usepackage{wallpaper} 
\usepackage{amsmath}
\usepackage{amsthm}
\usepackage{amsfonts}
\usepackage{subfigure}
\usepackage[latin1]{inputenc}
\usepackage{wasysym}
\usepackage{color}
\usepackage{xcolor}
\usepackage{xspace}
\usepackage{tikz}
\usepackage{cite}
\usepackage{mathtools}
\usepackage{graphicx}

\usepackage{lineno}

\usepackage{mathrsfs}

\newcommand{\Rfb}{R_{\text{FB}}\xspace}
\newcommand{\jpsi}{\ensuremath{{\text{J}/\psi}}\xspace}

\newcommand{\pT}{\ensuremath{p_{\text{T}}}\xspace}

\newcommand{\snn}{\ensuremath{\sqrt{s_{_{\text{NN}}}}}\xspace}

\newcommand{\pp}{\ensuremath{pp}\xspace}
\newcommand{\pPb}{\ensuremath{p\text{+Pb}}\xspace}
\newcommand{\PbPb}{\ensuremath{\text{Pb+Pb}}\xspace}

\newcommand{\avgNcoll}{\ensuremath{\langle N_{\text{coll}} \rangle}\xspace}

\newcommand{\W}{\ensuremath{W}\xspace}
\newcommand{\Z}{\ensuremath{Z}\xspace}

\newcommand{\MET}{\ensuremath{E_{\text{T}}^{\text{miss}}}\xspace}
\newcommand{\ET}{\ensuremath{E_{\text{T}}}\xspace}

\newcommand{\der}{\ensuremath{\text{d}}\xspace}

\newcommand*{\EtaL}{\ensuremath{\eta^{\ell}}\xspace}
\newcommand*{\EtaMu}{\ensuremath{\eta^{\mu}}\xspace}
\newcommand*{\EtaE}{\ensuremath{\eta^{e}}\xspace}
\title{Vector boson and charmonia measurements in \pPb collisions with ATLAS at the LHC}

\ShortTitle{Vector boson and Charmonia measurements in p+Pb collisions with ATLAS}

\author{\speaker{Markus K. K\"ohler}\thanks{On behalf of the ATLAS Collaboration}\\
        Weizmann Institute of Science\\
        E-mail: \email{markus.konrad.kohler@cern.ch}}

\abstract{The production of electroweak bosons ($Z,\gamma$ and $W$) and charmonia is sensitive to the initial-state 
geometry of heavy-ion collisions and to the parton distribution function with its potential nuclear 
modification. Since their leptonic decay products do not interact strongly, their kinematics are unmodified 
by the strongly interacting medium, which can be created in a heavy-ion collision. \\
We report on the latest results of the ATLAS Collaboration on electroweak boson and charmonia production in \pPb 
collisions at $\snn = 5.02$~TeV. Production yields of \Z and \W bosons are presented as a function of (pseudo-)rapidity 
in different centrality bins. The forward-backward ratio of $\jpsi$ is shown as a 
function of transverse momentum and center-of-mass rapidity.\\
}

\FullConference{XXIV International Workshop on Deep-Inelastic Scattering and Related Subjects\\
		11-15 April, 2016\\
		DESY Hamburg, Germany}

\begin{document}
%\linenumbers
\section{Introduction}
Heavy-ion collisions at ultrarelativistic energies, such as performed at the LHC at CERN, are believed to result in a 
phase transition from hadronic to partonic degrees of freedom and form a state of matter called Quark-Gluon 
Plasma (QGP), see e.g. Ref.~\cite{HI_review} for a recent review.\\
It is known from previous experiments that color charged particles are suppressed in heavy-ion 
collisions~\cite{ALICE_RAA}, which is attributed to the parton energy-loss by traversing through the hot 
and dense medium~\cite{E_Loss}. However, particles which are created in hard scatterings before the QGP is 
formed and whose decay products do not interact strongly provide the possibility to study heavy-ion 
collisions unmodified by the medium. So the production yield of electroweak bosons, such as prompt photons or \W and \Z 
bosons, can be described by a superposition of nucleon-nucleon collisions scaled by the mean number of collisions 
calculated within the Glauber model in \PbPb collisions at $\snn = 2.76$~TeV~\cite{ATLAS_PbPb_photons}. These 
measurements are consistent to NLO QCD calculations 
using a particle distribution function (PDF), which does not incorporate nuclear modifications. However, the 
measurements do not allow to exclude nuclear modifications within their precision.\\
To study PDFs and their potential nuclear modifications, it is advantageous to study asymmetric collision systems like 
\pPb collisions~\cite{PDF_pA}. Since nuclear modifications in \pPb collisions have a larger impact on the 
resulting 
boson distributions, measurements can better discriminate between PDFs which incorporate nuclear modifications and 
those which do not. \\
As opposed to electroweak bosons, heavy quarks, such as charm or beauty, do interact with the hot and dense matter. In 
central \PbPb collisions, \jpsi production is suppressed relative to \pp collisions~\cite{ALICE_JPsi}. To distinguish 
the impact of initial state and final state effects, charmonia production in \pPb collisions serves as a control 
measurement for \PbPb collisions.\\
These proceedings report on \W boson, \Z boson and charmonia measurements in \pPb collisions at $\snn = 5.02$~TeV 
collected by ATLAS\footnote{ATLAS uses a right-handed coordinate system with its origin at the nominal interaction 
point 
(IP) in the centre of the detector and the $z$-axis along the beam pipe. The $x$-axis points from the IP to the centre 
of the LHC ring, and the $y$-axis points upward. Cylindrical coordinates $(r,\phi)$ are used in the transverse plane, 
$\phi$ being the azimuthal angle around the $z$-axis. The pseudorapidity is defined in terms of the polar angle 
$\theta$ as $\eta=-\ln\tan(\theta/2)$.} \cite{ATL_det} at the LHC corresponding to an integrated luminosity of 
$28$~nb$^{-1}$. The Pb 
beam had an energy of $1.57$~TeV per nucleon and the opposing proton beam had an energy of $4$~TeV. As a convention, 
results are shown with positive rapidity corresponding to the proton beam direction and negative 
rapidity corresponding to the direction of the Pb beam, `Pb-going' side.
 
 \section{Electroweak bosons}
\begin{figure}[t]
\centering
 \includegraphics[scale = 0.22]{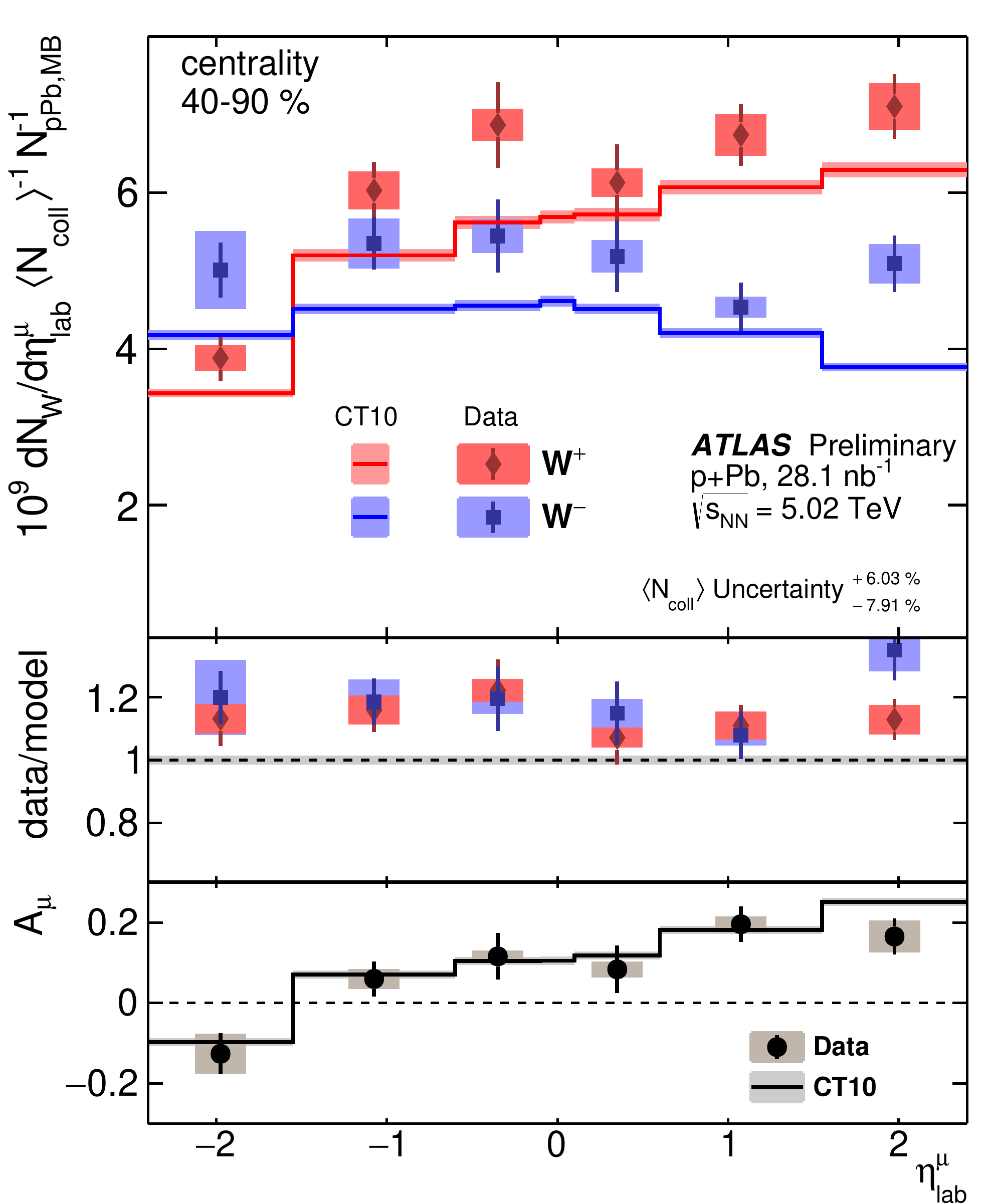}
 \hspace{0.5cm} \includegraphics[scale = 0.22]{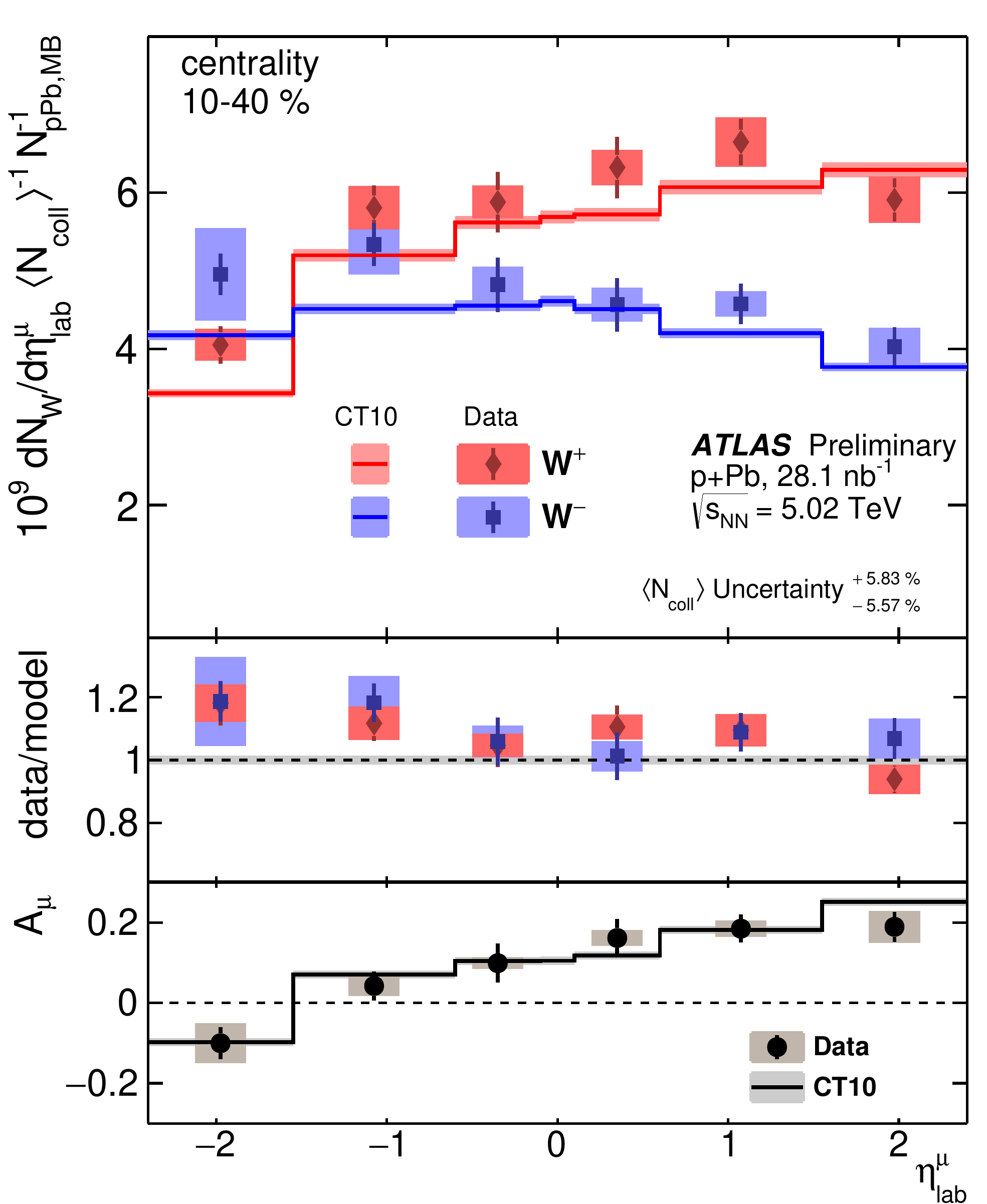}
 \hspace{0.5cm} \includegraphics[scale = 0.22]{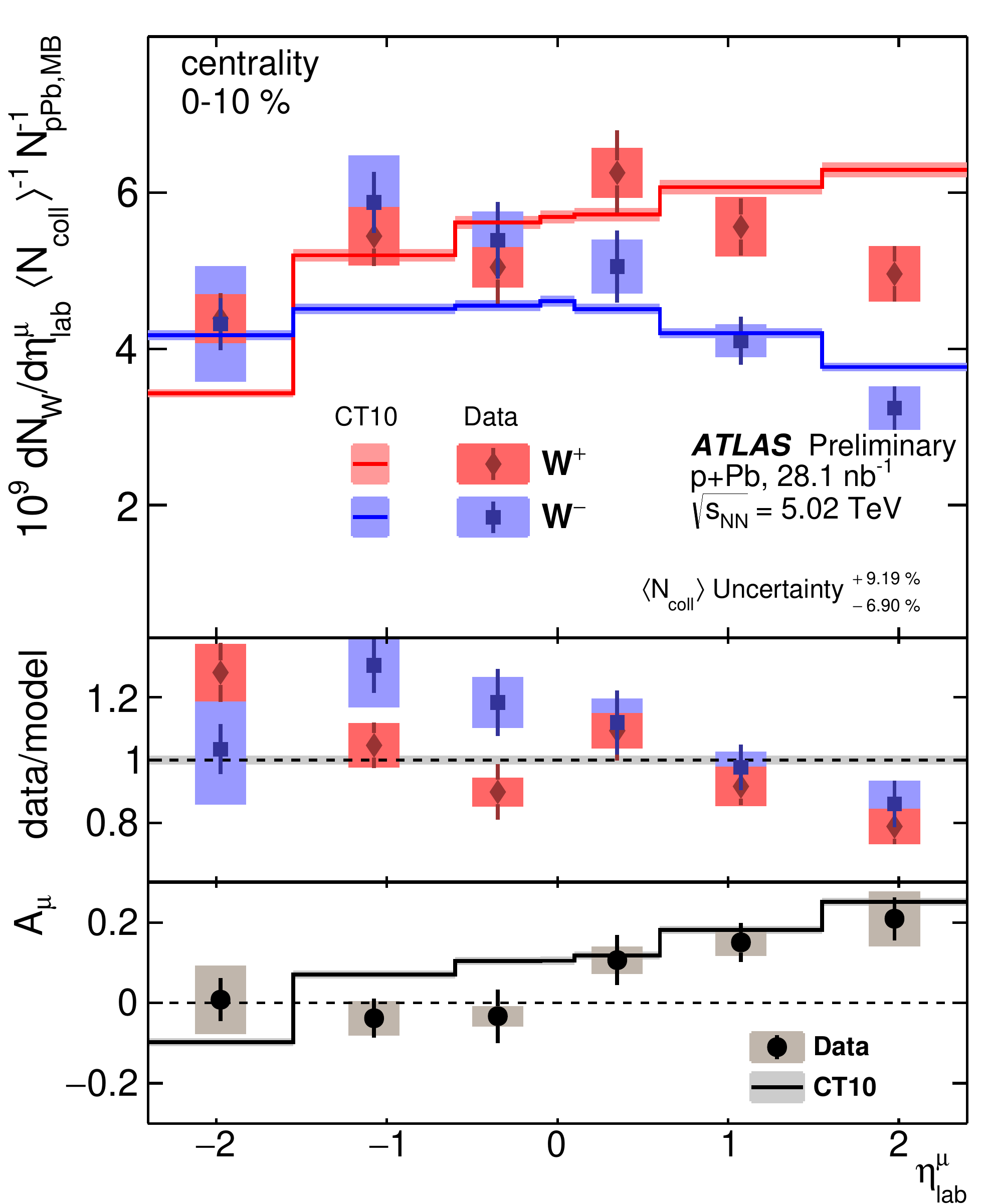}
 \caption{The \W boson production is shown as a function of the muon pseudo-rapidity in the laboratory frame for the 
centrality $40-90$~\% (left panel), $10-40$~\% (middle panel) and $0-10$~\% (right panel). In the 
upper panels, the production rate normalized to the number of minimum-bias events and the average number of binary 
nucleon-nucleon collision per centrality is shown. Data is compared to POWHEG calculations incorporating CT10. 
The ratio of data and calculations is shown in the middle panel. The lower panel shows the lepton charge 
asymmetry. Figures taken from~\cite{ATLAS_pPb_W}.} \label{fig:W_Eta_Cent}
\end{figure}

The \W boson production is measured in its muon decay channel. Muons are reconstructed in the Inner 
Detector and the Muon Spectrometer and combined using a $\chi^2$ minimization. The neutrino is identified by the 
signature of the missing transverse energy $\MET$. The production yields have been measured within the fiducial volume 
of $\pT^{\mu}> 25$~GeV, $\MET>25$~GeV, \mbox{$0.1 < \left|\EtaMu\right| < 2.4$} and $m_{\text{T}} = \sqrt{2 \pT^{\mu} 
\MET  (1- \cos\Delta \phi_{\mu,\MET})} > 40$~GeV, where  $\Delta \phi_{\mu,\MET}$ is the azimuthal angle between the 
muon and the missing transverse energy. The background, consisting of muons from multijet collisions and other 
electroweak muon sources, is estimated by fitting signal and background templates to the measured $\MET$ distribution.\\
\begin{figure}[t]
\centering
 \includegraphics[scale = 0.3]{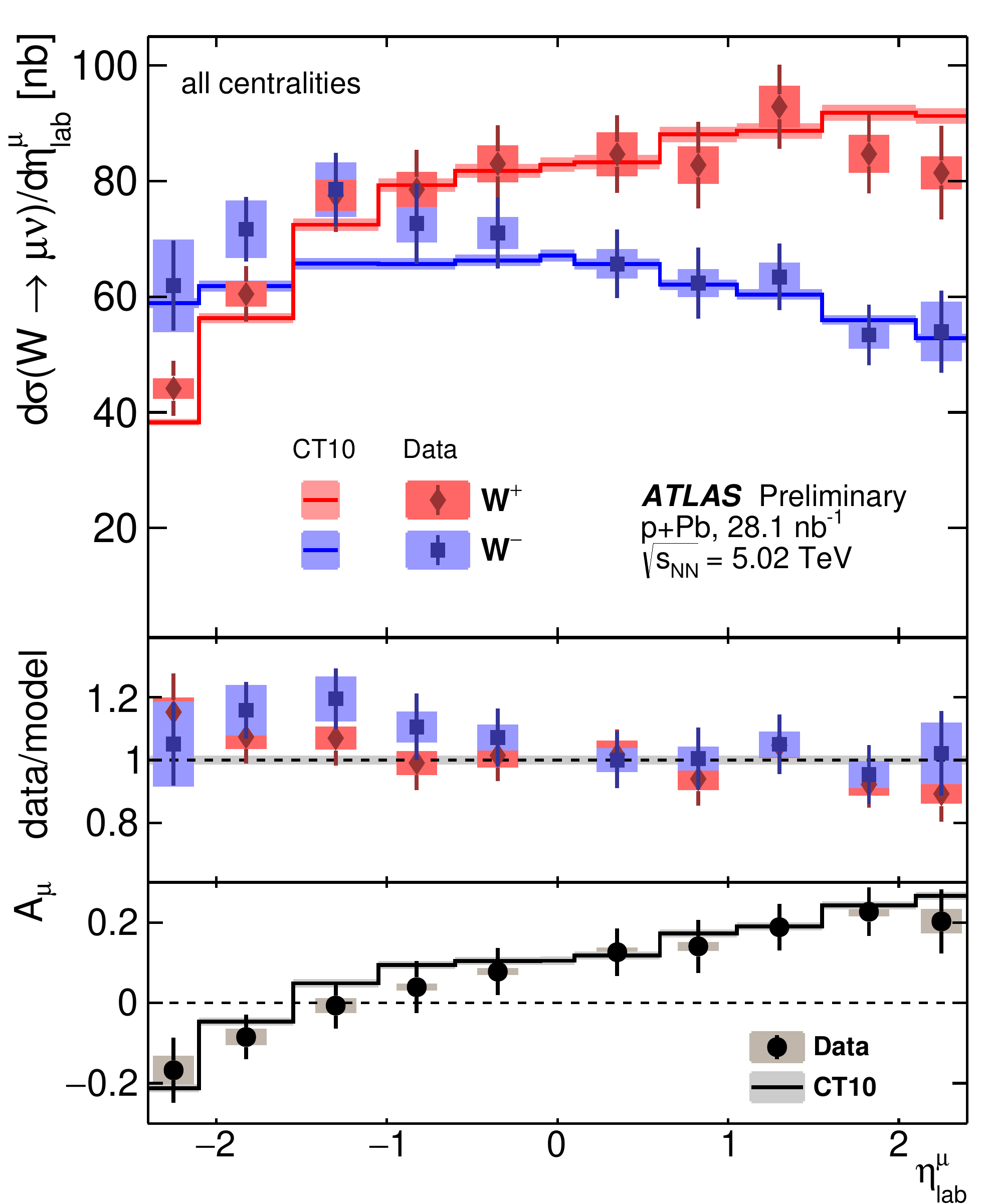} 
 \hspace{1cm} \includegraphics[scale = 0.3]{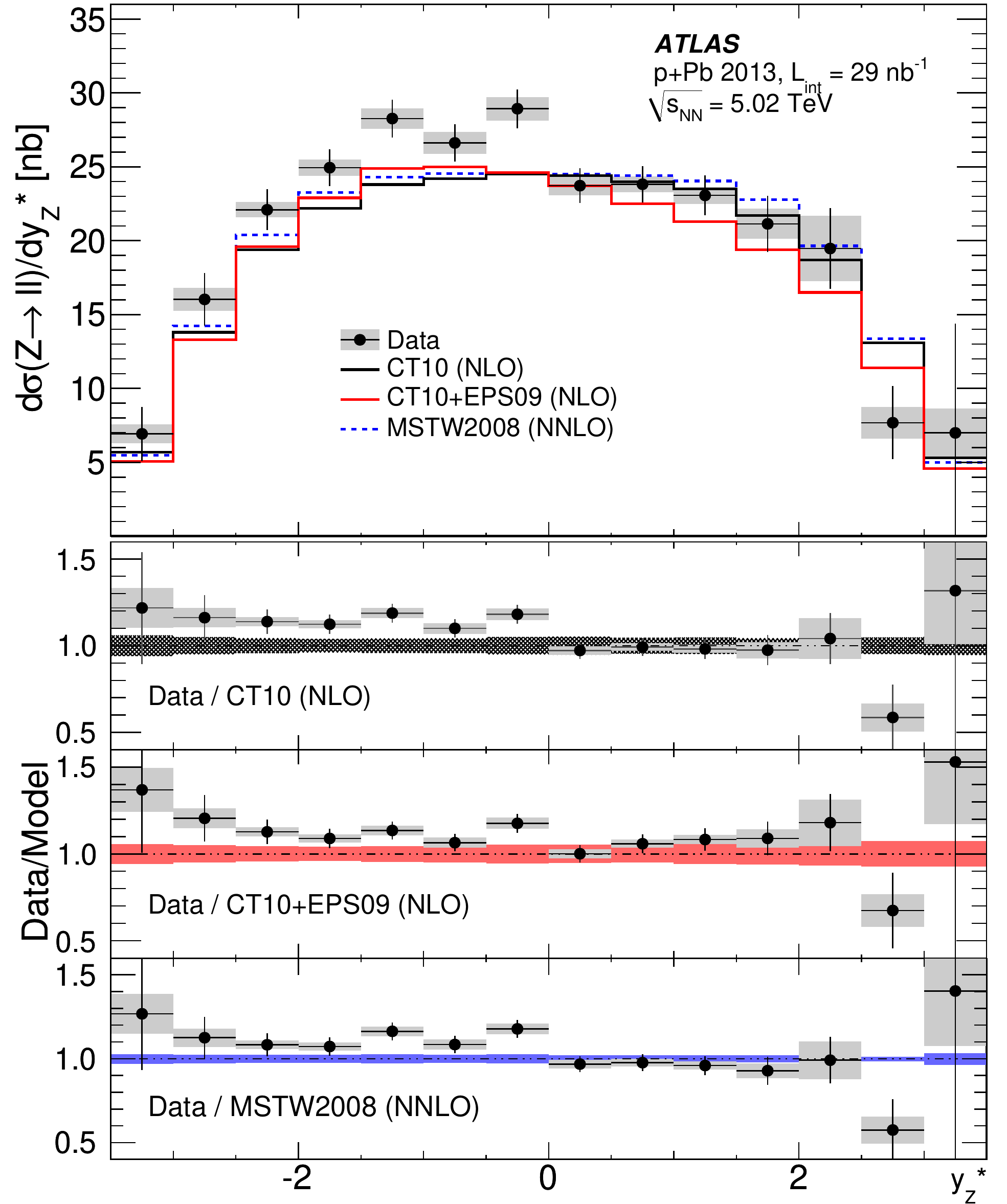}
 \caption{Cross section measurements of \W (left panel) and \Z (right panel) bosons in \pPb collisions at $\snn = 
5.02$~TeV as a function of 
(pseudo-) rapidity. Data is compared to NLO QCD calculations incorporating the PDF CT10. In addition, \Z boson 
production is also compared to MSTW2008 and to CT10 incorporating nuclear modifications EPS09. Figures taken 
from~\cite{ATLAS_pPb_W} and \cite{ATLAS_pPb_Z}.}\label{fig:XSec}
\end{figure}
The cross section is calculated separately for $W^+$ and $W^-$. The differences between both charges can be expressed 
by the lepton charge asymmetry $A_{\mu}$, and is given by
\begin{equation}
 A_{\mu} (\EtaL) = \frac{\der N_{W^+} / \der \EtaL - \der N_{W^-} /\der \EtaL } {\der N_{W^+}/\der \EtaL + 
\der N_{W^-}/ \der \EtaL},
\label{eq:ChAsymm}
\end{equation}
where $N_{W^+}$ and $N_{W^-}$ is the efficiency corrected and background subtracted number of $W^+$ and $W^-$.\\
\Z bosons were measured via their dielectron and dimuon decay channel. Electrons are measured within the 
pseudo-rapidity range $|\EtaE|<2.5$, excluding the transition region between the barrel and the endcap calorimeter 
$1.37 
< |\EtaE| < 1.52$. They are then required to have $\ET > 20$~GeV. Muons are analyzed within the fuducial volume $\pT > 
20$~GeV and $|\EtaMu| < 2.4$. The combinatorial background is estimated by the amount of like-sign pairs. After the 
dielectron and dimuon channel have been found consistent, both channels are combined with weights set by their 
corresponding uncertainties. \\
In Figure~\ref{fig:XSec}, the cross section of \W and \Z bosons is shown as a function of $\EtaMu$ and 
the center-of-mass rapidity\footnote{The center-of-mass rapidity $y^*$ is defined as $y^* = \frac{1}{2}\ln 
\frac{E+p_z }{E-p_z}$, where $E$ and $p_z$ are the energy and the component of the momentum along the proton beam 
direction in the nucleon-nucleon center-of-mass frame.} $y^*$. Data is compared to calculations using 
the PDF CT10~\cite{CT10} and (in the case of the \Z boson), 
MSTW2008~\cite{MSTW2008} and CT10 incorporating the nuclear modification EPS09~\cite{EPS09}. Additionally, the ratio 
between data and the calculations are shown. In the cases of the \Z and $W^-$ bosons, the measured cross sections 
exceed the calculations in the Pb-going side. Consequently, the lepton charge asymmetry undershoots the calculations in 
the Pb-going side. It should be noted that in the case of the \Z boson, the calculations of CT10 incorporating the 
nuclear modifications EPS09 described the data best~\cite{ATLAS_pPb_Z}. The same behavior for the cross section of 
electroweak bosons was also seen by the CMS Collaboration~\cite{CMS_pPb_W, CMS_pPb_Z}.\\
In addition to the cross sections, the centrality dependence of \W and \Z production is investigated. The values for  
\avgNcoll are calculated using the Glauber MC simulation~\cite{Glauber} using the total transverse energy 
deposited in the forward calorimeter (FCal) 
in the Pb-going direction. In addition, values for \avgNcoll are calculated using Glauber-Gribov Color Fluctuations 
(GGCF) models, see e.g.~\cite{GGCF}. \\
In presence of a hard scattering process, such as \W or \Z boson production, the underlying event is increased. 
Consequently, more energy is deposited in the FCal, leading to a potential bias of the centrality 
calculation. Therefore a correction was applied, following Ref.~\cite{Cent-Bias}.  \\
In Figure~\ref{fig:W_Eta_Cent}, the \W boson production as a function of $\EtaMu$ is shown for peripheral ($40 - 
90$~\%), mid-central ($10-40$~\%) and central ($0-10$~\%) collisions for the case of centrality calculations 
within the Glauber model and corrected for the centrality bias. The data shown is compared with calculations 
incorporating the 
CT10 PDF. Data is normalized to number of minimum-bias events and \avgNcoll within the centrality class.\\
In the most peripheral centrality class, data is consistently exceeding the calculations, which might be due the 
uncertainty in calculation of \avgNcoll. However, there appears to also be change of slope within the most central 
events. While in 
peripheral and mid-central collisions the lepton charge asymmetry is well described by the CT10 calculations, in the 
most central event class there is a discrepancy in differential cross section in the same rapidity range where it 
deviates from the model as shown in Figure~\ref{fig:XSec}.

\section{Charmonia}
\begin{figure}[t]
\centering
 \includegraphics[scale = 0.36]{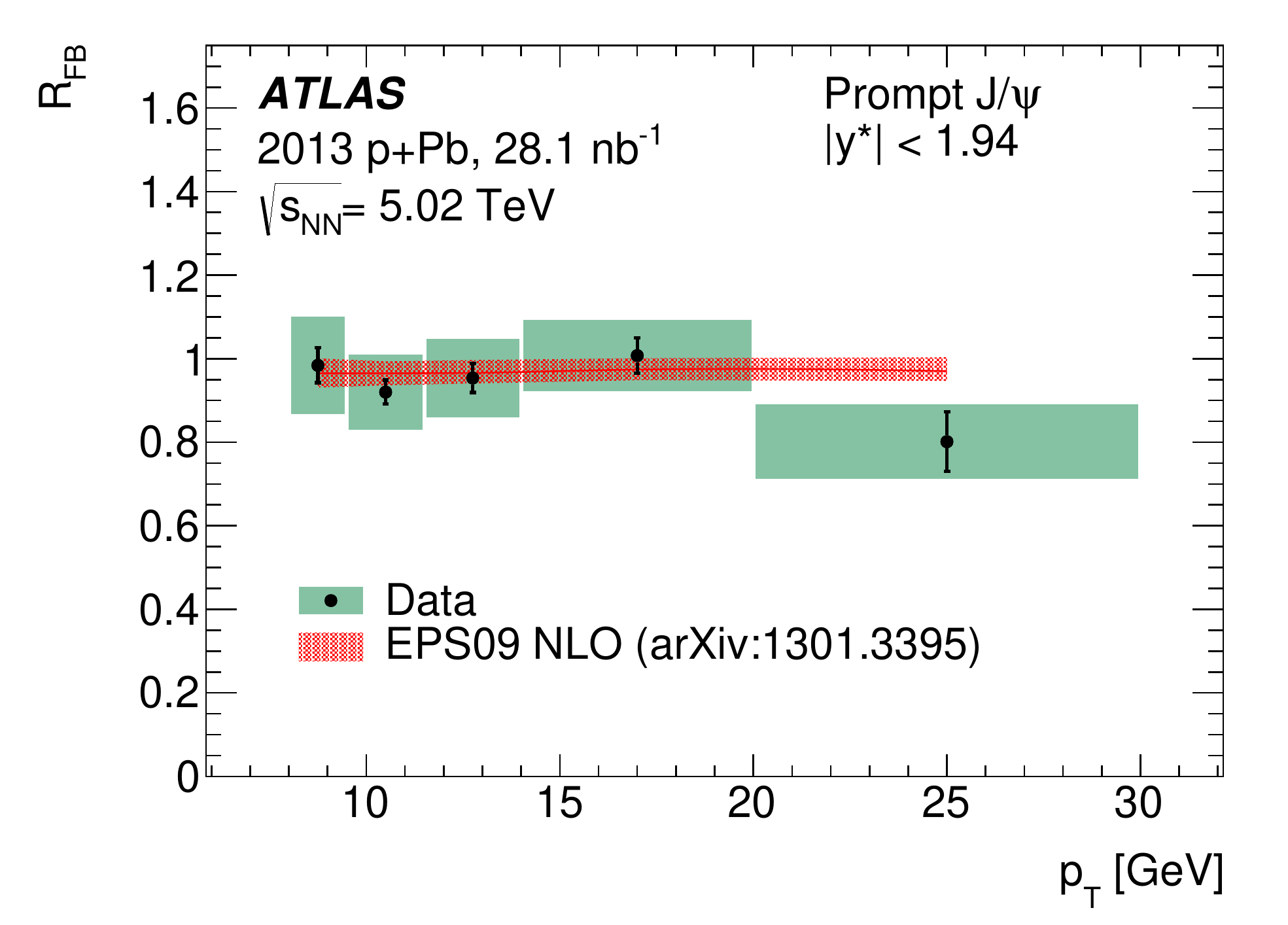}
 \includegraphics[scale = 0.36]{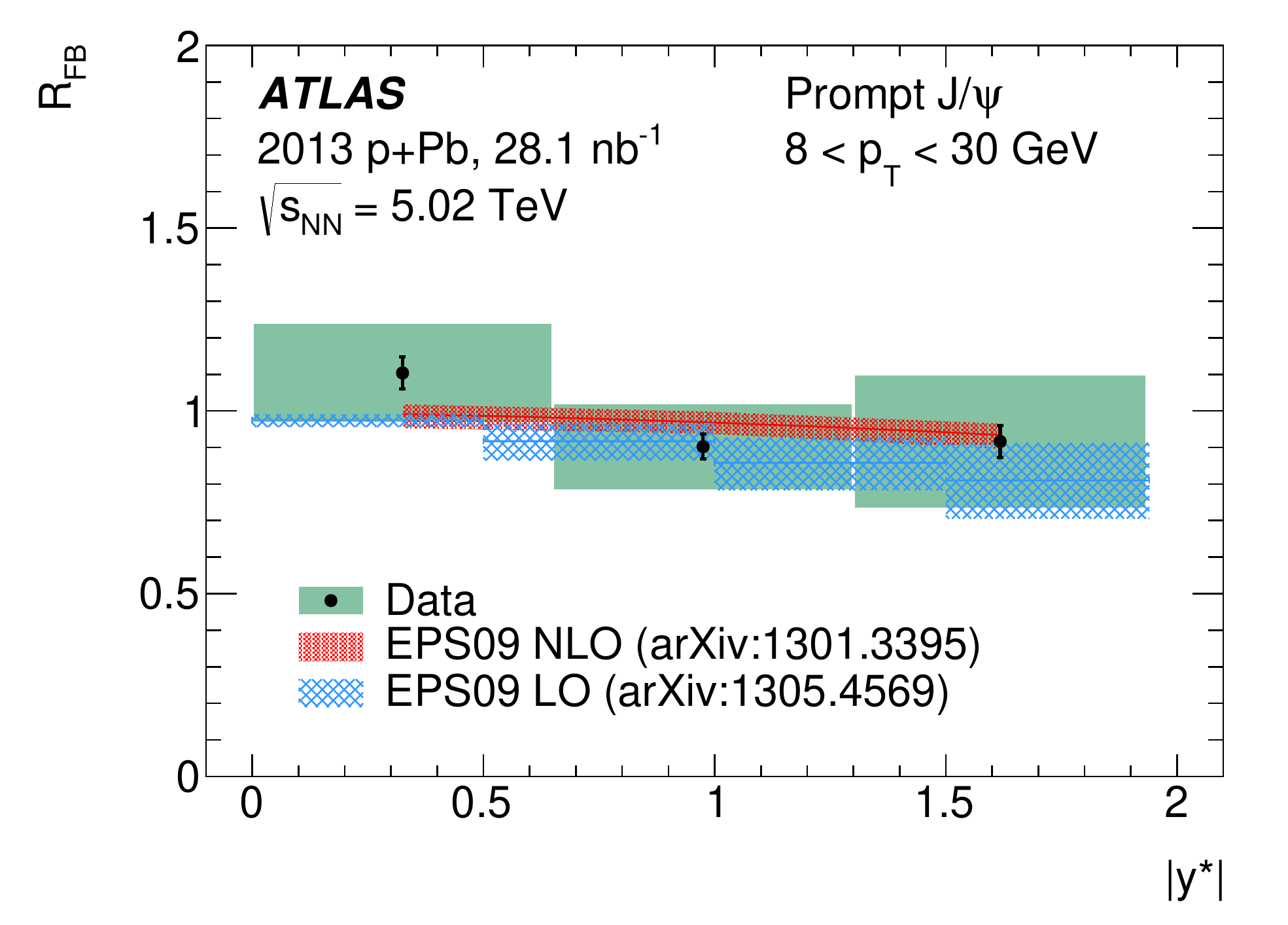}
 \caption{Forward-backward production ratio $\Rfb$ of prompt \jpsi as a function of \jpsi transverse momentum (left 
panel) and as a function of center-of-mass pseudo-rapidity (right panel). Figures taken from~\cite{ATLAS_pPb_JPsi}.} 
\label{fig:JPsi_FB}
\end{figure}

The measurement of \jpsi production in \pPb collisions was carried out in the dimuon channel. Muons are required to have 
$\pT > 4$~GeV and be within the acceptance of 
$|\eta|  < 2.4$. Non-prompt \jpsi contributions, i.e. \jpsi from decay chains of $b$-quarks, are separated by using 
the `pseudoproper time', $\tau = L_{xy} \frac{m_{\mu \mu}}{\pT}$, where $\pT$, $m_{\mu\mu}$ and $L_{xy}$ are the 
dimuon transverse momentum, its invariant mass and the signed transverse distance between the primary vertices and the 
$\jpsi$ decay vertex, respectively. \\
In asymmetric collision systems, such as $\pPb$ collisions, the cross sections might not be symmetric with respect 
to $y^* 
= 0$. These asymmetries of the cross sections can be quantified by the forward-backward production ratio $\Rfb$, which 
is defined as 
\begin{equation}
 \Rfb (\pT, y^*) = \frac{\der ^2 \sigma (\pT , y^* > 0) / \der \pT \der y^*}{\der ^2 \sigma (\pT , y^* < 0) / 
\der \pT \der y^*}.
\end{equation}
In Figure~\ref{fig:JPsi_FB}, $\Rfb$ of prompt $\jpsi$ is shown as a function of the \jpsi transverse momentum and the 
center-of-mass rapidity within $8 < \pT < 30$~GeV and $| y^* | < 1.94$. The experimental results are consistent with 
unity within the experimental uncertainties and no significant dependence on $\pT$ or $y^*$ is found. Data is compared 
with PDFs including nuclear modifications (EPS09) in LO and NLO. The calculations are found to be consistent with data.

\section{Summary}
We presented the latest results on \W boson, \Z boson and charmonia production in \pPb collisions at $\snn = 
5.02$~TeV measured with the ATLAS detector at LHC. \\
NLO pQCD calculations describe the \W and \Z boson production well, except in the Pb-going direction, where a small 
excess seems to appear. This behavior is best described by PDFs including nuclear modifications. \\
The forward-backward ratio of prompt \jpsi is shown and found to be consistent with unity within experimental 
uncertainties. This behavior is reproduced by two PDF sets incorporating EPS09 modifications.

\subsection*{Acknowledgements}
This research is supported by the Israel Science Foundation (grant 1065/15) and by the \mbox{MINERVA} Stiftung with the 
funds from the BMBF of the Federal Republic of Germany.

\end{document}